\documentclass[         %
superscriptaddress,    
nofootinbib,            
onecolumn,             %
floatfix]               
{revtex4-1}               

\usepackage{amssymb}
\usepackage{graphicx}
\usepackage{amsmath}
\usepackage[hypertex]{hyperref}
\usepackage{amsfonts}
\usepackage{natbib}
\usepackage{lipsum}

\usepackage{tabularx}
\usepackage{xcolor}
\usepackage{oubraces}
\usepackage{dsfont}
\usepackage{titlesec}
\usepackage{transparent}
\usepackage{tocloft}
\usepackage{color}

\begin{document}
\title{Proton tunneling in hydrogen
bonds \\
and its implications in an induced-fit model of enzyme
catalysis}
\author{Onur Pusuluk}
\affiliation{Department of Physics, \.{I}stanbul Technical
University, Maslak, \.{I}stanbul, 34469 Turkey}
\author{Tristan Farrow}
\affiliation{Department of Physics, University of Oxford, Parks
Road, Oxford, OX1 3PU, UK} \affiliation{Centre for Quantum
Technologies, National University of Singapore, 3 Science Drive 2,
Singapore 117543, Singapore}
\author{Cemsinan Deliduman}
\affiliation{Department of Physics, Mimar Sinan Fine Arts
University, Bomonti, \.{I}stanbul, 34380, Turkey}
\author{Keith Burnett}
\affiliation{University of Sheffield, Western Bank, Sheffield S10
2TN, UK}
\author{Vlatko Vedral}
\affiliation{Department of Physics, University of Oxford, Parks
Road, Oxford, OX1 3PU, UK} \affiliation{Centre for Quantum
Technologies, National University of Singapore, 3 Science Drive 2,
Singapore 117543, Singapore}

\date{\today}

\begin{abstract}

The role of proton tunneling in biological catalysis is investigated
here within the frameworks of quantum information theory and
thermodynamics. We consider the quantum correlations generated
through two hydrogen bonds between a substrate and a prototypical
enzyme that first catalyzes the tautomerization of the substrate to
move on to a subsequent catalysis, and discuss how the enzyme can
derive its catalytic potency from these correlations. In particular,
we show that classical changes induced in the binding site of the
enzyme spreads the quantum correlations among all of the four
hydrogen-bonded atoms thanks to the directionality of hydrogen
bonds. If the enzyme rapidly returns to its initial state after the
binding stage, the substrate ends in a new transition state
corresponding to a quantum superposition. Open quantum system
dynamics can then naturally drive the reaction in the forward
direction from the major tautomeric form to the minor tautomeric
form without needing any additional catalytic activity. We find that
in this scenario the enzyme lowers the activation energy so much
that there is no energy barrier left in the tautomerization, even if
the quantum correlations quickly decay.

\end{abstract}

\maketitle

\section{Introduction}

An enzyme is a macromolecule that catalyzes one or more specific
biological reactions without being consumed. Each reactant molecule
that an enzyme acts upon is known as a substrate of this enzyme. The
conversion of the substrate to one or more products involves the
formation of unstable intermediate structures called transition
states, while uncatalysed reactions can not occur at fast-enough
rates because of the high energies of these molecules. Enzymes
accelerate such slow-rate reactions by lowering the energy required
to form the transition state in several ways \cite{1999_Fersht,
2004_Science_ReviewOfES}, e.g. by destabilizing the substrate, by
stabilizing the transition state, or by leading the reaction into an
alternative chemical pathway.

Enzymes must first recognize their specific substrates before moving
on to catalyze the associated reactions. The favoured model for the
enzyme and the substrate interaction is the induced-fit mechanism
\cite{InducedFit_1958, InducedFit_2002}, which enhances the
recognition specificity in a noisy environment
\cite{InducedFit_2007}. This model states that although initial
intermolecular interactions are weak, they trigger a continuous
conformational change in the binding site of the enzyme. This
provides the structural complementarity between the enzyme and the
substrate, in the manner of a lock and key. The catalytic site of
the enzyme then accelerates the intermolecular conversion.

Another mainstream model for the enzyme - substrate interaction is
the conformational selection \cite{ConfSelect_1999,
ConfSelect_2002}. This model states that the enzyme exists in an
equilibrium between the active and inactive conformations until the
incoming substrate binding. Then, the intermolecular interaction
stabilizes the active conformation of the enzyme. Unlike the
induced-fit model, the conformational change upon the substrate
binding is not so significant in the conformational selection model.
However, it is still of importance in the catalytic activity of the
enzyme.

Enzymes are divided into six major classes depending on the type of
chemical reaction they catalyze: oxidoreductases, transferases,
hydrolases, lyases, isomerases, and ligases. The role of proton
tunneling in enzyme catalysis has been investigated so far
particularly focusing on the class of oxidoreductases that oxidize a
substrate by transferring its hydrogen (H) to an acceptor molecule
through intermolecular H-bonds.

According to the kinetic isotope effects studies, observed rates of
several enzyme-catalyzed oxidoreduction reactions cannot be
explained solely in terms of a semi-classical transition-state
theory without a quantum-mechanical correction for the tunneling of
H species \cite{HTInE_1989__KIE, HTInE_1996__KIE, HTInE_1998__KIE,
HTInE_1999_1__KIE_Conf, HTInE_1999_2__KIE_Conf, HTInE_2000__KIE,
HTInE_2009__KIE, HTInE_2010__KIE}. Moreover, a number of different
studies, including the kinetic isotope experiments
\cite{HTInE_1999_1__KIE_Conf, HTInE_1999_2__KIE_Conf}, molecular
dynamic simulations \cite{HTInE_1991__MD_Conf, HTInE_1997__MD_Conf,
HTInE_2001__MD_Conf}, quantum mechanical molecular mechanics
calculations \cite{HTInE_2006_1__QMMM_Conf, HTInE_2006_2__QMMM_Conf,
HTInE_2006_3__QMMM_Conf, HTInE_2012__QMMM_Conf,
HTInE_2013__QMMM_Conf}, and qualitative quantum rate models
\cite{Anders_2017_JPC_HTunnelingInSLO} give prominence to the role
of the conformational change in promoting tunneling.

Conversely, some experimental tests using artificial catalysts
\cite{NoEffectOfT_2003} and realistic simulations based on path
integral formulation \cite{NoEffectOfT_1991, NoEffectOfT_1996,
NoEffectOfT_2004, NoEffectOfT_2006, NoEffectOfT_2010, 2010_McKenzie}
suggest that quantum tunneling in some oxidoreductase (and also
lyase) enzymes does not enhance reaction rates sufficiently enough
to be regarded as catalytic. Hence, the role of tunneling of H
species in catalytic reactions is still open to debate
\cite{YesOrNo_2004, YesOrNo_2006, YesOrNo_2017}.

The debate summarized above generally focuses on the enzymatic
reactions that involve a proton transport between carbon atoms,
rather than the nitrogen or oxygen atoms that are expected to form
H-bonds strong enough to show a quantum character. Also, the
conformational changes of the enzymes are usually considered in the
context of conformational selection model rather than the
induced-fit model that requires a significant change in the
conformation. We extend the present debate to include strong H-bonds
and large conformational changes by taking into account the
tautomerization reactions in which protons are relocated inside the
substrate molecules.

A special subclass of structural isomerases known as the tautomerase
superfamily is normally responsible for the tautomerization
reactions. The reactions catalyzed by the known members of this
superfamily usually involve proton transport from and/or to a carbon
atom \cite{2008_TautomeraseSF} too. However, neither the nature of
tautomerization nor its role in biocatalysis is limited to
tautomerase activity. In particular, tautomerization of nucleotides
occurs generally via proton transfer between nitrogen and oxygen
atoms, and is also in charge of the initial stages of various
bio-catalyzed reactions.

Exchange of protons is likely between the carbonyl and/or amino
groups of nucleotides and solvent molecules. Thus, a single
nucleotide can exist in many tautomeric forms due to the
solvent-mediated tautomerization. One of these forms predominates
under physiological conditions and is called the major tautomer.
However, tautomeric preferences of different enzymes may be also
different, i.e., whereas the major tautomeric form of a nucleotide
is the substrate of an enzyme, another enzyme may require one of the
possible minor tautomeric forms of the same nucleotide. As an
example, to prevent point mutations, members of the transferase
class of enzymes that drive DNA replication should ensure that each
thymine nucleotide exists in its major tautomeric form during
catalytic activity. On the other hand, the DNA repair enzyme
Nei-like 1, a member of hydrolase class of enzymes, converts thymine
glycol nucleotides to minor tautomeric forms during the substrate
binding stage, and this tautomerization gives rise to a more
efficient catalytic activity \cite{2016_TautoRequiredByE}. Moreover,
nucleotide tautomerization also plays a direct role in a number of
different functions of RNA enzymes \cite{2015_TautomeraseRoleInRNA}.

In addition to the likelihood of quantum H-bonds in nucleotide
tautomerization and its importance in biological functions, the
induced-fit mechanism also appears to be common in nucleotide
recognition \cite{InducedFit_2013_DNA, InducedFit_2014_DNA,
InducedFit_2015_DNA-1, InducedFit_2015_DNA-2, InducedFit_2016_DNA}.

\section{Model and Methods} \label{Sec_Model}

We examine a generic molecular recognition event in which two
quantum H-bonds are formed between the substrate and a
multifunctional enzyme that requires the tautomerization of the
substrate to execute a different biological function which is
outside the scope of the paper. Consistent with the induced-fit
model, we allow a significant conformational change in the binding
site of the enzyme. Then, we approach the problem of proton motion
in these H-bonded system using the tools of quantum information
theory where correlations are routinely regarded as a resource for
specific tasks. Although the binding site is reshaped in a fully
classical way, we find that this classical motion increases the
quantum correlations in the intermolecular H-bonds and spreads them
among all of the four H-bonded atoms. Finally, we discuss the
possibility of using these quantum correlations as a resource for
the tautomerization of the substrate in the first catalytic stage of
multi-step enzyme catalysis.

\subsection{Biological scenario} \label{Sec_BioScen}
\begin{figure}[h]
    \centering
    \includegraphics[width=0.6 \linewidth]{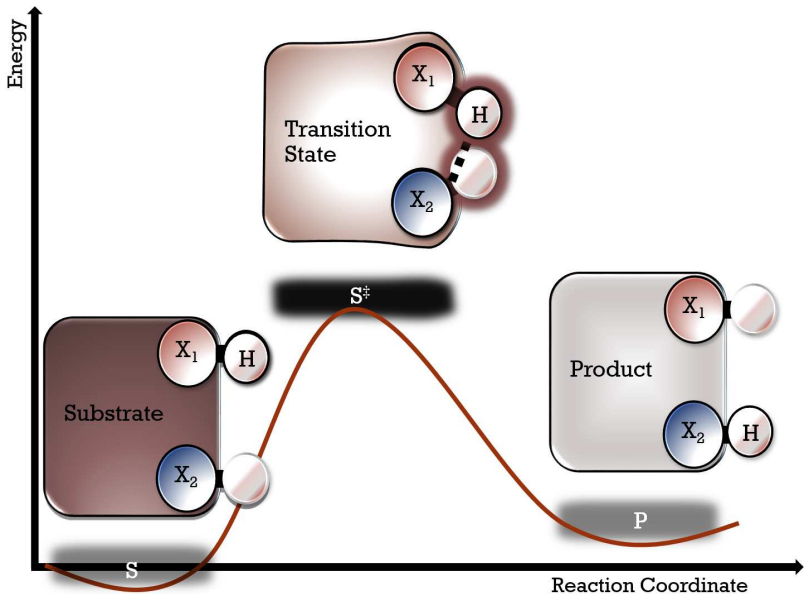}
    \caption{A substrate (S), transition state structure
    (S$^\ddagger$), and product (P) in a generic tautomerization reaction.
    S and P are constitutional isomers. Although the intermolecular conversion from S to P
    is nothing more than the movement of a proton from X$_1$ to
    X$_2$, a more reactive intermediate (S$^\ddagger$) is involved in the reaction.
    Since S$^\ddagger$ corresponds to
    a saddle point on the potential energy surface, the spontaneous tautomerization occurs
    very slowly.} \label{Fig_Substrat}
\end{figure}

We consider a generic nucleotide existing in its major tautomeric
form as the substrate (S) in a putative tautomerization event as
shown in Fig. \ref{Fig_Substrat}. Here, S is converted to a product
(P) that corresponds to a minor tautomeric form of the molecule
originating from the relocation of a proton from one electronegative
atom/group (X$_1$) to another (X$_2$) like oxygen or nitrogen. A
direct relocation due to the tunneling is not possible because of
the large bond angle $\phi_{12} \equiv \angle\text{HX}_1\text{X}_2$.
Thus, the molecule undergoes a conformational change resulting in an
unstable intermediate structure denoted by S$^\ddagger$. This new
conformation of the molecule allows a bond angle smaller than
$\pi/2$ that facilitates orbital interactions and proton tunneling
between X$_1$ and X$_2$. However, as S$^\ddagger$ corresponds to the
highest potential energy along the reaction coordinate, this
tautomerization reaction is not likely to occur on its own.

In aqueous solution, water molecules can mediate the interconversion
from S to P in a two-step reaction consisting of subsequent
protonation and deprotonation processes. If a proton-donating water
molecule interacts with X$_2$ first, a cation intermediate
S$^\ddagger_+$ precedes the P. Otherwise, the water-mediated
reaction involves the formation of an anion intermediate
S$^\ddagger_-$. However, P is expected to be rapidly converted back
to S in both cases. Hence, the probability of P in the equilibrium
should not be high enough for an enzyme that requires P to
initialize its catalytic function.
\begin{figure}[t]
    \centering
    \includegraphics[width=0.8\linewidth]{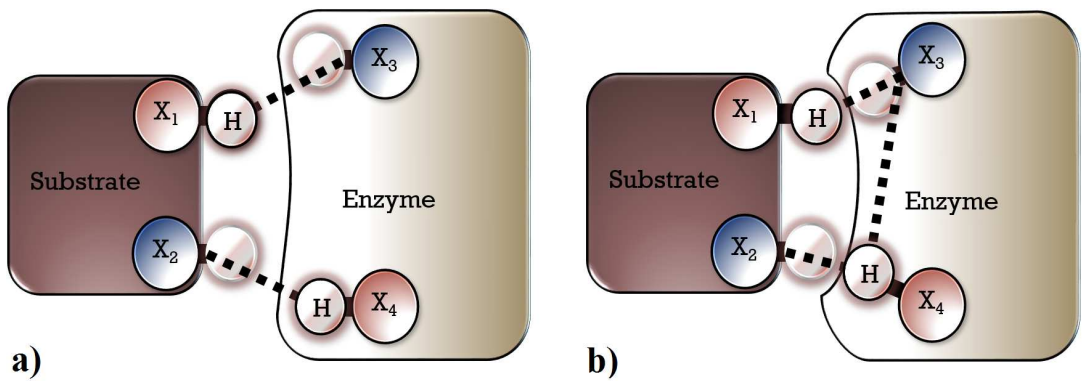}
    \caption{The generic induced-fit model under consideration.
    (a) Hypothetical interaction where two intermolecular
    H-bonds are formed between an enzyme and the substrate. (b) Changes
    occur in the binding site of the enzyme in accordance with the induced-fit
    model. Note that not only $\phi_{13}$ and $\phi_{42}$, but also $\phi_{43}$ becomes smaller.} \label{Fig_ES}
\end{figure}

Conversely, consider now the enzyme-catalyzed tautomerization during
a generic induced-fit recognition event where two H-bonds are formed
between the enzyme (E) and S (Fig. \ref{Fig_ES}). The
atoms/molecules X$_3$ and X$_4$ that participate in this
intermolecular interaction are continuously tilted by the
interaction until the bond angles $\phi_{13}$ and $\phi_{42}$ reach
to the values that maximize the strength of intermolecular H-bonds.
We assume here that (i) the binding site of E returns back to its
initial configuration at end of the first catalytic stage and (ii)
the following second catalytic stage cannot be initialized unless S
is converted to P before that moment.

We can comprehend time-scales associated with the conformational
transitions of the enzyme's binding site by using the following
analogy. We visualize an imaginary spring that connects the
unoccupied and the occupied proton locations near X$_3$ and X$_4$ in
the binding site (see Fig. \ref{Fig_ES}-a). The first conformational
change that occurs according to the induced-fit mechanism (Fig.
\ref{Fig_ES}-b) corresponds to the compression of the spring,
whereas the second conformational change that returns the enzyme
back to its native conformation corresponds to the extension of the
spring. Note that there is a close relationship between the first
conformational change and the binding energy of intermolecular
H-bonds: the initial binding energy of weak intermolecular H-bonds
triggers the first conformational change that in turn increases the
strength and the binding energy of H-bonds. Thus, the binding energy
should be visualized as an external force exerted on the spring that
is responsible for the compression of the spring from its
equilibrium length, as it increases with compression. However, the
deformation of the enzyme's conformation stops at the end of the
binding stage, which means that the compression of the spring should
come to an end at this point. This can be provided only by a
restoring force that applies to the spring and quantifies the
enzyme's tendency to return back to its initial state. This
restoring force should be initially smaller but increases faster
than the external force during compression. That is to say, the
spring under consideration is a nonlinear spring that hardens as it
is compressed. In this case, the compression of the spring
constantly decelerates and finally stops when the restoring force
becomes equal to the external force.

The enzyme - substrate complex does not become frozen after the
binding stage. Instead, the molecules decouple from each other and
the enzyme undergoes the second conformational change. This
corresponds to the point when the external force exerted on the
spring disappears and the spring immediately starts to extend to
reach its equilibrium length. The extension motion is rapid compared
to the preceding compression as the net force that drives the
process is not the difference between the external and restoring
forces but the restoring force itself. Thus, it is plausible to
describe the conformational change that decouples the binding site
of the enzyme from the substrate and converts it back to the initial
state on a time-scale $\tau_2$ that is much smaller than time-scale
$\tau_1$ over which the enzyme - substrate complex reaches the
optimum intermolecular binding energy.

\subsection{Physical model} \label{Sec_Phys}

\subsubsection{Motion of the proton between two atoms/molecules}
\begin{figure}[h]
    \centering
    \includegraphics[width=0.45 \linewidth]{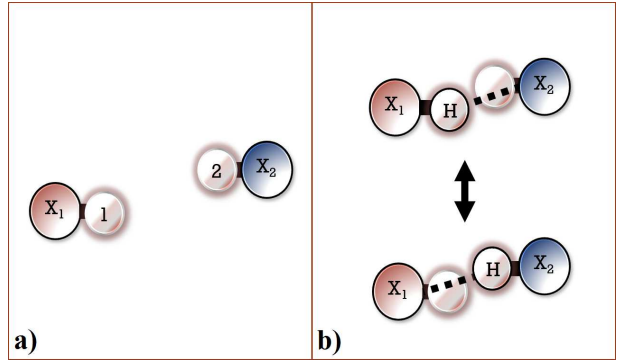}
    \caption{Definition of H-bonding.
    (a) A simple depiction of the H-bonded atoms/groups
    X$_j$. When the proton of the H atom resides at
    location $j = \{1, 2\}$, X$_j$ is called the proton-donor.
    (b) Elongation of the covalent bond between the H atom and
    the proton-donor (solid black stick) is an indicator of
    a H-bond (dashed black line). Due to intermolecular orbital
    interactions, the proton tunnels back and forth through
    the H-bond as shown by the two-sided arrow.}
\label{Fig_HBond}
\end{figure}

The locations between which the proton tunnels back and forth can be
regarded as the sites of an interaction network in such a way that
each site $j = \{1, 2\}$ is associated with the bonding orbital
$\sigma_{\text{X$_j-$H}}$ (Fig. \ref{Fig_HBond}-a). Then the proton
is expected to move from one site to another in accordance with the
Hamiltonian
\begin{equation}\label{Ham_HB1}
H_{H\!B} = \sum_{j=1}^{2} W_j \, \hat{n}_j - J_{12} (a_1^{\dagger}
a_{2} + a_1 a_{2}^{\dagger}) + V_{12} \, \hat{n}_1 \hat{n}_{2} +
\lambda_{12} \, \mathbb{I}_{12} ,
\end{equation}
where $\hat{n}_{j} = a_{j}^{\dagger} a_{j}$ is the proton number
operator at site $j$, $a_{j}^{\dagger}$ and $a_{j}$ are respectively
proton creation and annihilation operators that obey the fermion
anticommutation relations. On-site energy $W_{j}$ can be taken to be
the total potential felt by a proton at $j$th site. $J_{12}$ stands
for the orbital interactions that drive proton tunneling. $V_{12}$
is introduced to penalize the case in which there is one proton at
each site. $\lambda_{12}$ is a constant responsible for the total
intermolecular interactions between X$_1$ and X$_2$.

Each of the coefficients in Eq. (\ref{Ham_HB1}) exhibits a different
functional dependence on geometric parameters such as the length of
the single X$_j-$H covalent bond ($r_j$), the separation between
X$_1$ and X$_2$ ($R_{12}$) and the angle ($\phi_{12} \equiv
\angle\text{HX}_1\text{X}_2$). The key coefficient here is the
hopping constant $J_{12}$. Its functional dependence on the
geometric parameters can be written in a similar way to the coupling
constant of diabatic state models \cite{2012_McKenzie,
2014_McKenzie} as:
\begin{equation}\label{J}
J_{12} = J_0 \cos(\phi_{12}) \frac{R_{12} - r_1
\cos(\phi_{12})}{\sqrt{R_{12}^2 + r_1^2 - 2 R_{12} r_1
\cos(\phi_{12})}}\mathrm{e}^{-b_0 (R_{12} - R_0)} ,
\end{equation}
where $J_0$, $b_0$, and $R_0$ depend on the chemical identities of
X$_1$, X$_2$ and their environment. This functional dependence
guarantees the directionality that is a critical property of
H-bonds. $J_{12}$ becomes $J_0 \, \mathrm{e}^{-b_0 (R_{12} - R_0)}$
for linear H-bonds and decreases slightly by the increase of
$\phi_{12}$. When $\phi_{12}$ reaches to the value of $\pi / 2$, it
vanishes and the proton can not tunnel between X$_1$ and X$_2$
anymore even if the separation $R_{12}$ is sufficient for the
formation of a quantum H-bond. However, it is important to note that
our results in this study do not require the exact form given in
(\ref{J}). The only requirement is that $J_{12}$ should be non-zero
at small bond angles, and should vanish when the bond angle becomes
larger.

To obtain a pseudo-spin Hamiltonian by preserving the
anti-commutation relations, we apply the Jordan-Wigner
transformation to $a_{j}$, $a_{j}^{\dagger}$, and $\hat{n}_{j}$ in
(\ref{Ham_HB1}) with the convention for Pauli $z$ operator that
$\sigma_z^{(j)} = |0_{j}\rangle \langle0_{j}| - |1_{j}\rangle
\langle1_{j}|$. Then, we arrive at a two-spin Heisenberg XXZ model
under one homogeneous and one inhomogeneous magnetic field,
\begin{eqnarray} \begin{aligned} \label{Ham_HB2}
H_{H\!B} = J_x (\sigma^{(1)}_{x} \otimes \sigma^{(2)}_{x} +
\sigma^{(1)}_{y} \otimes \sigma^{(2)}_{y}) + J_z \, \sigma^{(1)}_{z}
\otimes \sigma^{(2)}_{z} + (B + b) \, \sigma^{(1)}_{z} + (B - b) \,
\sigma^{(2)}_{z} + \tilde{\lambda} ,
\end{aligned}
\end{eqnarray}
with parameters $J_x = - J_{12} / 2$, $J_z = V_{12} / 4$,
$\tilde{\lambda} = \lambda_{12} + (2 W_1 + 2 W_2 + V_{12}) / 4$, $B
= - (W_1 + W_2 + V_{12}) / 4$, $ b = - (W_1 - W_2) / 4$. The
eigensystem of this pseudo-spin Hamiltonian can be written in an
increasing order of eigenvalues as:
\begin{eqnarray}
\begin{cases}
e_1 = - J_z - 2 \sqrt{b^2 + J_{x}^2} + \tilde{\lambda}, & \mbox{}
|e_1\rangle \mbox{} = (\eta_- |01\rangle + |10\rangle) / (1 +
\eta_-^2)^{\frac{1}{2}} , \\
e_2 = - J_z + 2 \sqrt{b^2 + J_{x}^2} + \tilde{\lambda}, & \mbox{}
|e_2\rangle \mbox{} = (\eta_+ |01\rangle + |10\rangle) / (1 +
\eta_+^2)^{\frac{1}{2}} , \\
e_3 = - 2 B + J_z + \tilde{\lambda}, & \mbox{} |e_3\rangle \mbox{} =
|11\rangle , \\
e_4 = + 2 B + J_z + \tilde{\lambda}, & \mbox{} |e_4\rangle \mbox{} =
|00\rangle ,
\end{cases}
\end{eqnarray}
where $\eta_{\pm} = (b \pm \sqrt{b^2 + J_{x}^2})/J_{x}$. The ground
state $|e_1\rangle$ is just a two-qubit representation of the state
$\alpha |\text{X$_1-$H}\cdot\cdot\cdot\!\text{X$_2$}\rangle + \beta
|\text{X$_1$}\!\!\cdot\cdot\cdot\text{H$-$X$_2$}\rangle$ that
describes a generic H-bond. Unlike the one-qubit representation of
the same state that is used in the spin-boson and the diabatic
states models found in the literature, this two-qubit representation
can investigate quantum correlations such as entanglement and
discord generated in H-bonds. This is an advantage of our quantum
informational perspective that stands to yield new insights since
there is a high amount of correlations in biomolecular interactions
and quantum physics is better suited to describe all correlations
contained in a complex system in a rigorous way
\cite{Anders_2011_Chaos_ComplexityAndquantum}. Also, the ground
state energy $e_1$ goes to the expected energy of a classical bond,
i.e. $W_1 + \lambda$, when $J_{x}$ vanishes. Moreover, this energy
seems to be suitable for energy decomposition analysis as it equals
$W_1 + \lambda + J_{12}$ for symmetric H-bonds.

Besides these properties of the ground state, the last two excited
states respectively represent the state of X$_1$ and X$_2$ atoms of
the ionic intermediate structures S$^\ddagger_+$ and S$^\ddagger_-$
formed in water-mediated tautomerization.

In the following, each of the locations allowed for the protons is
regarded as a pseudo-spin, explained above. The formation of a
H-bond between two atoms/molecules $X_j$ and $X_{j^\prime}$ in the
biological scenario under consideration is then described as the
coupling of the corresponding pseudo-spins by the Hamiltonian $H^{(j
j^\prime)}_{H\!B}$ given in (\ref{Ham_HB2}).

\subsubsection{Tautomeric transitions due to the motion of protons}

Here, we will represent S, P, and all the possible transition states
between them by different eigenstates of a single Hamiltonian. This
may seem like an oversimplification, since each of these different
molecules is actually corresponding to the ground state of a
separate Hamiltonian with many electronic levels. However, it is
just a projection of all the extrema of the potential energy
surfaces of the nucleotide onto a single energy spectrum,
representing a unification of all the tautomeric forms along
possible reaction coordinates from S to P into a hypothetical
generic molecule. This type of modeling, which assigns a global
time-independent Hamiltonian for a transition from an initial to a
final state is quite common in quantum thermodynamics.

In this respect, we collect all the other degrees of freedom apart
from the two pseudo-spins in the molecular structure into a single
macromolecular configuration $C$ and label its state as
$|\zeta\rangle_C$. That is to say, the whole state $|N\rangle$
representing the nucleotide's tautomeric form $\text{N} =
\{\text{S}, \text{P}, \text{S}^{\ddagger},
\text{S}^{\ddagger}_{-},\text{S}^{\ddagger}_{+}, ...\}$ is taken to
be the product of its configuration state $|\zeta_\text{N}\rangle_C$
and its pseudo-spin state $|\psi_\text{N}\rangle_{12}$.

Let's first consider the states of S and P. To prevent a direct
transition between these forms, both $\phi_{12}$ in S and
$\phi_{21}$ in P are taken to be $\pi/2$. Then the hopping constant
$J_{12}$ substituted in $H^{(12)}_{H\!B}$ that describes the
pseudo-spin interaction in either form vanishes, i.e.
$|\psi_\text{S}\rangle_{12} = |10\rangle_{12}$ and
$|\psi_\text{P}\rangle_{12} = |01\rangle_{12}$. In other words, the
H-bonding interaction between X$_1$ and X$_2$ does not have a
quantum character. When one of the forms is converted to the other,
the energy of this classical interaction also changes. However, no
significant difference in the molecular structure is expected due to
this change. Thus, we can assign the same configuration state
$|G\rangle_C$ to both forms: $|S\rangle = |G\rangle_C \otimes
|10\rangle_{12}$ and $|P\rangle = |G\rangle_C \otimes
|01\rangle_{12}$.

Similarly, neither the protonation of X$_2$ nor the deprotonation of
X$_1$ is likely to alter the rest of the molecular structure. Hence,
the ionic transition states S$^\ddagger_-$ and S$^\ddagger_+$ can
respectively be described by $|G\rangle_C \otimes |00\rangle_{12}$
and $|G\rangle_C \otimes |11\rangle_{12}$. Conversely, a more
energetic configuration, $|E\rangle_C$, should be assigned to the
neutral transition state S$^\ddagger$ as the proton tunneling
between X$_1$ and X$_2$ requires a bond angle less than $\pi/2$ as
is given Fig. \ref{Fig_Substrat}. Then we end up with
$|S^\ddagger\rangle = |E\rangle_C \otimes |e_1\rangle_{12}$.

Note that the ground and excited states of the molecular
configuration, $|G\rangle_C$ and $|E\rangle_C$, which we can
visualize them like macromolecular logic qubits $|0\rangle_L$ and
$|1\rangle_L$, need not to include any quantum degree of freedom.

At this point, it is possible to describe tautomerization processes
as transitions between energy levels of a single Hamiltonian
constructed as:
\begin{eqnarray} \begin{aligned} \label{Ham_S}
H_{N} = E_g |G\rangle_C \langle G| &+ E_e |E\rangle_C \langle E| \\
&+ |G\rangle_C \langle G| \otimes H_{H\!B}^{(12)}(r,R,\phi = \pi/2)
\\ &+ |E\rangle_C \langle E| \otimes
H_{H\!B}^{(12)}(r^{\prime},R^{\prime},\phi^{\prime} < \pi/2) ,
\end{aligned}
\end{eqnarray}
with an eigensystem:
\begin{eqnarray} \label{ES_S}
\begin{cases}
\epsilon_1 = E_g + W_1 + \lambda_{12} , & \mbox{} |\epsilon_1\rangle
\mbox{} = |G\rangle_C \otimes |10\rangle_{12} \equiv |S\rangle ,
\\
\epsilon_2 = E_g + W_2 + \lambda_{12} , & \mbox{} |\epsilon_2\rangle
\mbox{} = |G\rangle_C \otimes |01\rangle_{12} \equiv |P\rangle ,
\\
\epsilon_3 = E_e + \epsilon_- , & \mbox{} |\epsilon_3\rangle \mbox{}
= |E\rangle_C \otimes |\epsilon_-\rangle_{12} \equiv
|S^{\ddagger}\rangle ,
\\
\epsilon_4 = E_e + \epsilon_+, & \mbox{} |\epsilon_4\rangle \mbox{}
= |E\rangle_C \otimes |\epsilon_+\rangle_{12} \equiv
|\tilde{S}^{\ddagger}\rangle ,
\\
\epsilon_5 = E_g + \lambda_{12} , & \mbox{} |\epsilon_5\rangle
\mbox{} = |G\rangle_C \otimes |00\rangle_{12} \equiv
|S^{\ddagger}_-\rangle ,
\\
\epsilon_6 = E_g + V_{12} + \lambda_{12} , & \mbox{}
|\epsilon_6\rangle \mbox{} = |G\rangle_C \otimes |11\rangle_{12}
\equiv |S^{\ddagger}_+\rangle ,
\\
\epsilon_7 = E_e + \lambda^\prime_{12}, & \mbox{} |\epsilon_7\rangle
\mbox{} = |E\rangle_C \otimes |00\rangle_{12} \equiv
|\tilde{S}^{\ddagger}_-\rangle ,
\\
\epsilon_8 = E_e + V^\prime_{12} + \lambda^\prime_{12}, & \mbox{}
|\epsilon_8\rangle \mbox{} = |E\rangle_C \otimes |11\rangle_{12}
\equiv |\tilde{S}^{\ddagger}_+\rangle ,
\end{cases}
\end{eqnarray}
where $\epsilon_- = e_1(r^{\prime},R^{\prime},\phi^{\prime} <
\pi/2)$ and $\epsilon_+ = e_2(r^{\prime},R^{\prime},\phi^{\prime} <
\pi/2)$. Also, $\tilde{\text{S}}^{\ddagger}$,
$\tilde{\text{S}}^{\ddagger}_-$, and $\tilde{\text{S}}^{\ddagger}_+$
are the first possible excitations from S$^\ddagger$, S$^\ddagger_-$
and S$^\ddagger_+$ respectively.

\subsubsection{Open quantum system dynamics}
\label{Sec_OQS}

The configuration of atoms/groups and the pseudo-spins are expected
to be coupled to different phonon environments in different ways.
For example, the configuration states are likely to exchange energy
with a heat bath $B$ in a reversible manner
\begin{eqnarray}\label{Ham_BC}
H_{BC} = \sum_k g_{k} \, |G\rangle_C \langle E| \otimes
b_{k}^{\dagger} + g^*_{k} \, |E\rangle_C \langle G| \otimes b_{k} ,
\end{eqnarray}
where $b_{k}^{\dagger}$ and $b_{k}$ are phonon creation and
annihilation operators associated with the $k$th oscillator mode of
the bath. These bath operators can be related to the rotational
vibration modes such as bending or libration modes that change the
orientation of X$_1$ and X$_2$. Such rotational vibrations may occur
due to the collisions with solvent molecules or as a result of the
intramolecular nucleotide dynamics. As X$_1$ and X$_2$ are
covalently bonded to the rest of the molecule, these vibrations
should be dependent on the orientation of the whole other
atoms/groups close to them.

However, dominant environmental effect on the proton should be
originated from the X$_j-$H stretch vibrations that aren't expected
to affect rest of the molecule in a significant way. Charge
fluctuations in the surrounding molecules may drive these
oscillations and we can incorporate them into our model by coupling
the position of the proton linearly to the equilibrium positions of
phonons through
\begin{eqnarray}\label{Ham_BS}
H_{\tilde{B}S} = \sum_{j = \{1,2\}} \sigma_z^{(j)} \otimes \sum_k
\tilde{g}_{k,j} \, (\tilde{b}_{k,j}^{\dagger} + \tilde{b}_{k,j}) .
\end{eqnarray}
where $\tilde{b}_{k,j}^{\dagger}$ and $\tilde{b}_{k,j}$ are phonon
creation and annihilation operators of a second heat bath
$\tilde{B}$ and they are associated with the $k$th oscillator mode
at the $j$th proton location. $H_{\tilde{B}S}$ guarantees that the
interaction of the pseudo-spins with this second heat bath
$\tilde{B}$ destroys the quantum correlations between them and has
no further effect on the configuration.

As the magnitudes of $E_g$ and $E_e$ are significantly higher than
the other components in the eigenenergies given in Eq. (\ref{ES_S}),
the evolution of configuration and pseudo-spins in the nucleotide
can be assumed to be separated in such a way that the former is not
affected by the latter, but not vice versa. In this respect, we can
solve a master equation for the configuration and use this solution
in the evolution of the pseudo-spins. Here, the evolution of either
the pseudo-spin or the configuration state is described using the
Markovian master equation in the Lindblad form
\cite{BreuerAndPetruccione-2002}, which is one of the key elements
of the theory of quantum thermodynamics
\cite{Kosloff_2013_MasterEqsInQTD},
\begin{eqnarray}\label{MasterEq}
\frac{d\rho}{dt} = - \frac{\text{i}}{\hbar} [H + \hbar
H_{L\!S},\rho] + \mathcal{D}(\rho) ,
\end{eqnarray}
where $H_{L\!S}$ is the Lamb shift Hamiltonian providing an
environment-induced unitary contribution to the dynamics:
\begin{eqnarray}\label{LambShift}
H_{L\!S} = \sum_{\omega} \sum_{j, j^{\prime}} S_{j
j^{\prime}}(\omega) \, A_{j}^{\dagger}(\omega)\,
A_{j^{\prime}}(\omega) ,
\end{eqnarray}
whereas $\mathcal{D}$ is the dissipator responsible for the
irreversible dynamics:
\begin{eqnarray} \begin{aligned} \label{Dissip}
\mathcal{D}(\rho) = \sum_{\omega} \sum_{j,j^{\prime}} \gamma_{j
j^{\prime}}(\omega) \, \big(A_{j^{\prime}}(\omega) \rho
A_{j}^{\dagger}(\omega) - \frac{1}{2} \{A_{j}^{\dagger}(\omega)\,
A_{j^{\prime}}(\omega), \rho \} \big) . \end{aligned}
\end{eqnarray}

While solving this master equation for the pseudo-spin state,
$\omega$ and $A_{j}(\omega)$ are respectively taken to be the Bohr
frequencies and the Schr\"{o}dinger picture eigenoperators of the
pseudo-spin Hamiltonian $H_{H\!B}$ in (\ref{Ham_HB2}), i.e., $\omega
= (e_{j^{\prime}} - e_{j^{\prime \prime}})/\hbar$ and $A_{j}(\omega)
= \sum_{j^\prime \! j^{\prime \prime}} |e_{j^{\prime \prime}}\rangle
\langle e_{j^{\prime \prime}}| A_{j} |e_{j^{\prime}}\rangle \langle
e_{j^{\prime}}|$ where $A_{j}$ are the Pauli $z$ operators in
accordance with the Eq. (\ref{Ham_BS}). Also, the coefficients $S_{j
j^{\prime}}(\omega)$ and $\gamma_{j j^{\prime}}(\omega)$ are defined
as the imaginary part and one half of the real part of the one-sided
Fourier transformation of the thermal correlation function of heat
bath $\tilde{B}$.

Parameters of the pseudo-spin Hamiltonian $H_{H\!B}$ used in this
solution should change depending on the configuration state in line
with the Eq. (\ref{Ham_S}). To do so, if the configuration is in the
state $|G\rangle_C$, $\omega$ and $A_{j}(\omega)$ are evaluated
using the eigensystem of $H_{H\!B}$ that is calculated with a
vanishing hopping constant $J_{x}$. This makes the dynamics of the
pseudo-spins fully separable from each other and corresponds to a
pure two-qubit dephasing process. Conversely, a non-vanishing
$J_{x}$ is taken into account when the configuration is in the state
$|E\rangle_C$. This couples the dynamics of the pseudo-spins to each
other and leads to an evolution that brings the pseudo-spin states
to a detailed balance only between $|e_1\rangle_{12}$ and
$|e_2\rangle_{12}$ in the stationary state.

On the other hand, the evolution of configuration state should be
independent from the pseudo-spin state. Hence, the master equation
(\ref{MasterEq}) is solved for the configuration state using the
Bohr frequencies and eigenoperators of an effective Hamiltonian $H_C
\approx E_g |G\rangle_C \langle G| + E_e |E\rangle_C \langle E|$.
However, the eigenoperators $A_{j}(\omega)$ cannot be directly
constructed from $H_C$ and $H_{BC}$. The interaction Hamiltonian
(\ref{Ham_BC}) is first decomposed into the form $\sum_j A^{(C)}_j
\otimes B^{(B)}_j$ with $A^{(C)}_j$ and $B^{(B)}_j$ are Hermitian
operators on the Hilbert spaces of the configuration $C$ and the
bath $B$. Moreover, the coefficients $S_{j j^{\prime}}(\omega)$ and
$\gamma_{j j^{\prime}}(\omega)$ are defined for the heat bath $B$
this time. This solution describes a thermalization process.

\subsubsection{Enzyme's conformational changes during the
induced-fit mechanism} \label{Sec_TimeScales}

Two allowed proton locations in the binding site of E are also
regarded as pseudo-spins, e.g., $|\psi_\text{E}(t_0)\rangle_{34} =
|01\rangle_{34}$. Furthermore, we take into account the conformation
of this site in a similar way and denote it by $C^\prime$:
$|\zeta_\text{E}(t_0)\rangle_{C^\prime} =
|G^\prime\rangle_{C^\prime}$. However, unlike the nucleotide's
conformation $C$, $C^\prime$ is assumed to change in a classical and
gradual way because of the intermolecular interaction so that its
pseudo-spins are continuously tilted. Thus, the bond angles
$\phi_{13}$ and $\phi_{42}$ become smaller and the resultant
decrease in the energy of intermolecular H-bonds strengthens the
binding interaction.

In accordance with the biological scenario under consideration (see
Sec. \ref{Sec_BioScen}), the time-scale $\tau_1$ of the binding
interaction due to the induced-fit mechanism is supposed to be much
larger than $\tau_2$, the time-scale over which E undergoes a
conformational change that converts $C^\prime$ back to the initial
state $|G^\prime\rangle_{C^\prime}$.

As the interaction Hamiltonian changes slowly in time relative to
the conformational change that follows, the instantaneous
eigenstates of the binding interaction are assumed to evolve
independently. Hence, the enzyme - substrate complex (ES) is
considered to be in the corresponding ground state at every instant
of time until it reaches the maximum complementarity between E and
S. On the other hand, as E rapidly returns back to its initial state
after this point due to the relation $\tau_2 \ll \tau_1$, the
reverse conformational change is supposed to be driven by a sudden
post-selection measurement on the pseudo-spins of E which
accompanies the loss of energy from its configuration $C^{\prime}$
to the heat bath $B$.

\subsection{Model Parameters}

Bath descriptions in our model can be made to include realistic
correlation functions that allow us to probe the actual dynamics in
real-time. It is also possible to work in the equilibrium without
specifying the bath correlation functions. In fact, as will be shown
in the following section, the steady state of the chosen master
equation requires the extraction of only two parameters, $E_e - E_g$
and $\epsilon_+ - \epsilon_-$ from the biochemical data.

$E_e - E_g$ can be estimated using the activation energy values of
the nucleotide tautomerization processes found in the literature. As
an example, the energy barrier for the intramolecular single proton
transfer on the Watson-Crick edge is calculated in the gas phase as
$41.90$ and $47.42$ kcal/mol for thymine \cite{Parameters_T}, $34.3$
kcal/mol for guanine \cite{Parameters_G}, and $45.6$ kcal/mol for
adenine\cite{Parameters_A}. To be consistent with these values, we
set $E_e - E_g$ to $42$ kcal/mol when required in what follows.

Since $\epsilon_+ - \epsilon_-$ equals to $\big((W_1 - W_2)^2 + 4
J_{12}^2\big)^{1/2}$, we write $\epsilon_+ - \epsilon_- \geq W_2 -
W_1 = \epsilon_2 - \epsilon_1$. That is to say, the value of
$\epsilon_+ - \epsilon_-$ is  bounded below by the energy difference
between the S and P tautomers of the nucleotide. This difference is
calculated as $13.08$ and $18.75$ kcal/mol for thymine
\cite{Parameters_T}, $1.3$ kcal/mol for guanine \cite{Parameters_G},
and $12.6$ kcal/mol for adenine\cite{Parameters_A} in the
tautomerization processes given in the previous paragraph. Hence,
the ratio of $(E_e - E_g)/(\epsilon_2 - \epsilon_1)$ is equal to
$3.20336$ and $2.52907$ for thymine, $26.3846$ for guanine, and
$3.61905$ for adenine. Excluding the guanine because of the apparent
incompatibility with the others, the mean of this ratio is found to
be $3.11716$. Then we can substitute $1/3 \times (E_e - E_g) = 14$
kcal/mol for $\epsilon_+ - \epsilon_-$ since it should be just above
the value of $\epsilon_2 - \epsilon_1$ for relatively small values
of $J_{12}$.

Unless stated otherwise, we generate numerical predictions from the
model using the values given above. To explore the sensitivity of
the predictions to these values, we squeeze the energy levels of the
Hamiltonian given in (\ref{Ham_S}) when it is necessary. In this
respect, we decrease the value of $E_e - E_g$ down to $34$ kcal/mol,
while increasing the value of $\epsilon_+ - \epsilon_-$ up to $19$
kcal/mol.

\newpage

\section{Results and Discussions}

\subsection{Uncatalysed
Reaction}

In the absence of enzymes that catalyze the reaction, the
interaction between the nucleotide and two heat baths $B$ and
$\tilde{B}$ is responsible for the spontaneous tautomerization
$|S\rangle \leftrightarrow |S^{\ddagger}\rangle \rightarrow
|P\rangle$. Such a tautomerization can occur only if each of the
interactions $H_N$, $H_{B C}$, and $H_{\tilde{B}S}$ comes to the
fore one by one in the order given in what follows: the
initialization of the transformation from $|S\rangle$ to $|P\rangle$
requires the excitation of configuration $C$ due to an energy
absorbtion from the heat bath $B$ through $H_{BC}$. The resultant
state $|E\rangle_C \otimes |10\rangle_{12}$ is a coherent
superposition of $|\epsilon_3\rangle = |S^{\ddagger}\rangle$ and
$|\epsilon_4\rangle$. Immediately after this excitation, the
self-Hamiltonian of the molecule drives the evolution towards
S$^{\ddagger}$ as $\epsilon_3 < \epsilon_4$. The transformation from
this transition state to $|P\rangle$ depends on (i) the destruction
of the quantum correlations of the pseudo-spins through
$H_{\tilde{B}S}$ and (ii) the loss of energy from the configuration
atoms/groups through $H_{BC}$. These two interactions in arbitrary
order lead to a mixture of $|S\rangle$ and $|P\rangle$.

As the configuration state represents the whole atoms/groups in
nucleotide (except for the proton of the H atom whose locations are
regarded as pseudo-spins), the occurrence of its excitation is
expected to be quite rare. Since the spontaneous inter-conversion of
tautomers requires this excitation to be initialized, its occurrence
is expected to be rare also. We show this mathematically here.

The master equation approach introduced above relies on the
assumption that the configuration energy $E_g$ or $E_e$ dominates
the total energy of the nucleotide (see Sec. \ref{Sec_OQS}). This
allows us to first investigate the configuration state dynamics
alone and then explore the pseudo-spin state dynamics for the
predetermined configurations in time.

The initial state in an uncatalysed reaction is $|S\rangle =
|G\rangle_C \otimes |10\rangle_{12}$. The steady-state solution of
the master equation of the configuration state does not depend on
the initial state and equals $P_g |G\rangle \langle G|_C + P_e
|E\rangle \langle E|_C$ where $P_{g/e} = e^{- \beta E_{g/e}} / (e^{-
\beta E_g} + e^{- \beta E_e})$. When the configuration state is
fixed to $|G\rangle_C$, there isn't any coupling between the
pseudo-spins and so the interaction $H_{BC}$ leads to decoherence.
Hence, the initial pseudo-spin state $|\psi(t_0)\rangle_{12} =
|10\rangle_{12}$ that does not carry any quantum coherence remains
the same during the open system dynamics described by the Eq.
(\ref{MasterEq}). On the other hand, there is a non-vanishing
coupling between the pseudo-spins when the configuration state is
fixed to $|E\rangle_C$. As long as the initial pseudo-spin state
$|\psi(t_0)\rangle_{12}$ lives in a subspace spanned by
$\{|01\rangle, |10\rangle\}$, $H_{BC}$ drives the evolution to a
detailed balance between $|\epsilon_-\rangle_{12}$ and
$|\epsilon_+\rangle_{12}$ according to the chosen master equation,
i.e., the pseudo-spins' state relaxes to the athermal attractor
state $\rho_{12}^{eq} = P_- |\epsilon_-\rangle \langle\epsilon_-| +
P_+ |\epsilon_+\rangle \langle\epsilon_+|$ where $P_\pm =  e^{-
\beta \epsilon_\pm} / (e^{- \beta \epsilon_+} + e^{- \beta
\epsilon_-})$. Hence, the open system dynamics bring the state
$|S\rangle$ towards
\begin{eqnarray} \begin{aligned}
\rho_{\text{N}}^{eq} &= P_g |G\rangle \langle G|_C \otimes
|10\rangle \langle10|_{12} + P_e |E\rangle \langle E|_C \otimes
\rho_{12}^{eq} \\
&= P_g |S\rangle \langle S| + P_e P_- |S^\ddagger\rangle \langle
S^\ddagger| + P_e P_+ |\tilde{S}^\ddagger\rangle \langle
\tilde{S}^\ddagger| ,
\end{aligned} \end{eqnarray}
in the stationary state. It means that we will observe the molecule
in the substrate state with a probability of $p(S) = P_g$ at
equilibrium. This probability is approximately equal to unity at
physiological temperatures for the choice of $E_e - E_g = 42$
kcal/mol. Similarly, the equilibrium probability of the occurrence
of transition state becomes $p(S^\ddagger) = P_e P_-$. This
probability is equal to $2.84 \times 10^{-30}$ for $E_e - E_g = 42$
kcal/mol and $\epsilon_+ - \epsilon_- = 14$ kcal/mol at $T = 37.5
^\circ C$. Increasing the temperature enhances it slightly but never
exceeds the order of $10^{-30}$ at physiological temperatures. On
the other hand, the molecule cannot be converted to the product
state in this approximation, even if the environment reaches
extremely high temperatures as the final state
$\rho_{\text{N}}^{eq}$ is orthogonal to the $|P\rangle$, i.e., $p(P)
= tr\big[\rho_{\text{N}}^{eq} \, |P\rangle \langle P|\big] = 0$.

\subsection{Water-Mediated Reaction}

Before moving on to the role of quantum H-bonds in enzyme-catalyzed
tautomerization, we investigate first this role in water-mediated
tautomerization. To do so, we take into account the formation of two
H-bonds between the nucleotide and one or more water molecules.
Allowed proton locations on the opposite sides of pseudo-spins 1 and
2 are also regarded as pseudo-spins and respectively labeled by
$\bar{1}$ and $\bar{2}$.

We assume that the intermolecular H-bonds described by
$H_{H\!B}^{(1\bar{1})}$ and $H_{H\!B}^{(2\bar{2})}$ generate quantum
correlations between the nucleotide and the water molecule(s) as:
\begin{equation}
|\psi\rangle_{1\bar{1}2\bar{2}} = (\alpha |10\rangle +
\sqrt{1-\alpha^2} |01\rangle)_{1\bar{1}} \otimes (\beta |01\rangle +
\sqrt{1-\beta^2} |10\rangle)_{2\bar{2}} .
\end{equation}

We consider the open system dynamics just after the generation of
these intermolecular correlations. First, we eliminate the water's
degrees of freedom $\bar{1}$ and $\bar{2}$ by taking a partial trace
over them and and focus on the reduced system dynamics of the
nucleotide. In this case, the initial state of the nucleotide is an
incoherent superposition state:
\begin{equation}
|\psi(t_0)\rangle_{\text{N}} = |G\rangle_C \otimes \{\alpha^2
\beta^2, |10\rangle; (1-\alpha^2) \beta^2, |00\rangle; \alpha^2
(1-\beta^2), |11\rangle; (1-\alpha^2) (1-\beta^2), |01\rangle
\}_{12} .
\end{equation}

Open system dynamics of this incoherent superposition ends up in the
following state:
\begin{eqnarray} \begin{aligned}
\rho_{\text{N}|\text{H}_2\text{O}}^{eq} &= \alpha^2 \beta^2 P_g
|S\rangle\langle S| + (1-\alpha^2) (1-\beta^2) P_g |P\rangle\langle
P|
\\
& \quad + (1-\alpha^2) \beta^2 P_g |S^\ddagger_-\rangle\langle
S^\ddagger_-| + \alpha^2 (1-\beta^2) P_g |S^\ddagger_+\rangle\langle
S^\ddagger_+| \\
& \quad + (1-\alpha^2) \beta^2 P_e
|\tilde{S}^\ddagger_-\rangle\langle \tilde{S}^\ddagger_-| + \alpha^2
(1-\beta^2) P_e |\tilde{S}^\ddagger_+\rangle\langle
\tilde{S}^\ddagger_+| \\
& \quad + (1 - \alpha^2 - \beta^2) P_e \big(P_-
|S^\ddagger\rangle\langle S^\ddagger| + P_+
|\tilde{S}^\ddagger\rangle\langle \tilde{S}^\ddagger|\big) .
\end{aligned} \end{eqnarray}

If the intermolecular H-bonds generate maximal entanglement between
the nucleotide and the water molecule(s), i.e., $\alpha = \beta =
1/\sqrt{2}$, the nucleotide reaches an equilibrium such that $p(S) =
p(S^\ddagger_-) = p(S^\ddagger_+) = p(P) = P_g/4$ that equals to
$0.25$ for the chosen model parameters. The balance between these
four tautomers can be driven through S or P by changing the values
of $\alpha$ and $\beta$ corresponding to a decrease in the amount of
total entanglement generated through the H-bonds.

The values of $\alpha$ and $\beta$ depend on the geometric
parameters of the interaction which is totally random, i.e.,
although the tautomeric conversion of S to P may be mediated by
means of the quantumness of water - nucleotide H-bonds, there is a
very little chance for the molecules in aqueous solution to show the
right orientations that provide the required values of $\alpha$ and
$\beta$.

\subsection{Enzyme Catalyzed Reaction}

Before the interaction, the nucleotide and the enzyme both exist in
their ground states:
\begin{eqnarray} \begin{aligned}
|\psi(t_0)\rangle_{\text{NE}} &= |S\rangle \otimes |E\rangle \\
&= |G\rangle_C |10\rangle_{12} \otimes
|G^{\prime}\rangle_{C^{\prime}} |01\rangle_{34} .
\end{aligned} \end{eqnarray}

The pseudo-spin term of this state can be thought as the ground
state of the interaction Hamiltonian $H_I = H_{H\!B}^{(13)} +
H_{H\!B}^{(24)}$ with large pseudo-spin separations and vanishing
hopping constants. When $R_{13}$ and $R_{24}$ become sufficiently
small, a small amount of entanglement is generated through the weak
X$_1-$H$\cdot\cdot\cdot$X$_3$ and X$_2$$\cdot\cdot\cdot$H$-$X$_4$
bonds:
\begin{eqnarray} \begin{aligned}
|\psi(t_1)\rangle_{\text{NE}} &= |G\rangle_C \otimes
|G^{\prime}\rangle_{C^{\prime}} \otimes (\alpha |10\rangle +
\sqrt{1-\alpha^2} |01\rangle)_{13} \otimes (\beta |01\rangle +
\sqrt{1-\beta^2} |10\rangle)_{24} .
\end{aligned} \end{eqnarray}

This weak interaction then induces a classical and gradual
conformational change in the binding site of E according to the
induced-fit mechanism and the bond angles $\phi_{13}$ and
$\phi_{42}$ become smaller. As explained in Sec.s \ref{Sec_BioScen}
and \ref{Sec_TimeScales}, this process occurs on a relatively large
time-scale $\tau_1$ and the pseudo-spins stay in the ground state of
the time-dependent interaction Hamiltonian $H_I$ at every instant of
time. Hence, this conformational motion not only strengthens the
binding interaction in ES complex due to a decrease in the energy,
it also increases the quantum correlations of the intermolecular
H-bonds.

In the meantime, the angle $\phi_{43}$ and the inter-spin separation
$R_{43}$ both change during the binding stage (see Fig.
\ref{Fig_ES}). Hence, the intramolecular interaction between X$_3$
and X$_4$ gains a quantum character as well. In this respect, the
total pseudo-spin interaction Hamiltonian $H_I(t)$ should be taken
as $H_{H\!B}^{(13)}+H_{H\!B}^{(24)}+H_{H\!B}^{(34)}$ when $t_2 >
t_1$.

When the interaction-induced conformational change stops at time
$t_3 > t_2$, the binding site of the enzyme should end up in its
highest excited state $|E^{\prime}\rangle$, whereas the pseudo-spins
should be in the ground state of $H_I(t_3)$:
\begin{eqnarray} \begin{aligned}
|\psi(t_3)\rangle_{\text{NE}} &= |G\rangle_C \otimes
|E^{\prime}\rangle_{C^{\prime}} \\
&\quad \otimes  (a |0011\rangle + b |0101\rangle + c |0110\rangle +
d
|1001\rangle + e |1010\rangle + f |1100\rangle)_{1234} \\
&\equiv |ES\rangle.
\end{aligned} \end{eqnarray}

It follows from the state of the reaction intermediate ES that the
entanglement generated at the end of the binding stage is shared
among all of the four pseudo-spins. Next, we show that the
four-qubit entanglement generated in this way can be transferred to
S when the configuration and pseudo-spins of E subsequently return
their initial states, which can enable the conversion of S to P. We
will also consider the case in which the four-qubit entanglement of
the reaction intermediate ES decays rapidly before the next
conformational change of the enzyme.

\subsubsection{Quantum correlated H-bonds}

We assume without any loss of generality that the quantum
correlations of each three H-bonds in Fig. \ref{Fig_ES}-b is
sustained until E undergoes a conformational change that converts it
back to the initial state. In other words, $\tau_2$ is assumed to be
small compared to the decoherence time $\tau_D$ enforced by the heat
bath $\tilde{B}$. This is quite reasonable as the H-bonded
atoms/groups are partially isolated from their environment until the
detachment of the ES complex.

As explained in Sec.s \ref{Sec_BioScen} and \ref{Sec_TimeScales},
the second conformation motion of interest drives $C^\prime$ from
$|E^\prime\rangle$ to $|G^\prime\rangle$ and is governed by the
interaction Hamiltonian $H_{BC^{\prime}}$. During the course of this
relatively fast motion, the pseudo-spins of E should detach from the
pseudo-spins of S and return back to the initial product state
$|01\rangle_{34}$. We describe this conformation-induced process by
a sudden post-selection measurement $\mathbf{M} = \mathbf{I}_{12}
\otimes |01\rangle_{34} \langle01|$ and find that some of the
four-qubit entanglement of the reaction intermediate ES can be
transferred to the pseudo-spins of S when $\mathbf{M}$ converts the
pseudo-spins of E back to their initial state:
\begin{eqnarray} \begin{aligned} \label{State_ESin1}
\mathbf{M} + H_{BC^{\prime}}: |ES\rangle \rightarrow
|\psi(t_4)\rangle_{\text{NE}} &= |G\rangle_C \otimes (b^\prime
|01\rangle +
d^\prime |10\rangle)_{12} \otimes |E\rangle \\
&= \big(b^\prime |P\rangle + d^\prime |S\rangle\big) \otimes
|E\rangle \\
&\equiv |ES^\ddagger\rangle ,
\end{aligned} \end{eqnarray}
where $b^\prime = b/\sqrt{b^2+d^2}$ and $d^\prime = \sqrt{1 -
{b^\prime}^2}$. We can analyze the open system dynamics of the
nucleotide using the master equation approach introduced in Sec.
\ref{Sec_OQS}, starting at $t = t_4$. The stationary solution
becomes
\begin{eqnarray} \begin{aligned} \label{State_ESfin1}
\rho_{\text{N}|qH\!B}^{eq} &= {d^{\prime}}^2 P_g |G\rangle \langle
G|_C \otimes |10\rangle \langle10|_{12} + {b^{\prime}}^2 P_g
|G\rangle \langle G|_C \otimes |01\rangle \langle01|_{12} + P_e
|E\rangle \langle E|_C \otimes
\rho_{12}^{eq} \\
&= {d^{\prime}}^2 P_g |S\rangle\langle S| + {b^{\prime}}^2 P_g
|P\rangle\langle P| + P_e P_- |S^\ddagger\rangle\langle S^\ddagger|
+ P_e P_+ |\tilde{S}^\ddagger\rangle\langle \tilde{S}^\ddagger| .
\end{aligned} \end{eqnarray}

This means that the enzyme converts the nucleotide into the product
state with a probability of $p(P) = {b^{\prime}}^2 P_g$ at
equilibrium. Also, neither $p(S^\ddagger)$ nor
$p(\tilde{S}^\ddagger)$ changes when compared to equilibrium
probabilities in the uncatalysed case.

$b^{\prime}$ depends on the strength of the H-bonds formed in the
enzyme-substrate interaction, as does the efficiency of the
catalysis. If the three H-bonds shown in Fig. \ref{Fig_ES}-b
generate a maximal W-type four-qubit entanglement, i.e., $a = b = c
= d = e = f = 1/\sqrt{6}$, the nucleotide reaches an equilibrium
such that $p(S) = p(P) = P_g/2$ that equals $0.5$ for our chosen
model parameters.

We can compare this result with another obtained in water-mediated
tautomerization. When the amounts of entanglement generated through
each water - nucleotide H-bond are chosen to be maximal, the
nucleotide reaches an equilibrium with $p(S) = p(P) = 0.25$. Here,
when the overall four-qubit entanglement is chosen to be maximal,
the pairwise entanglement generated through each enzyme - nucleotide
H-bond is not maximal and the equilibrium probabilities
corresponding to both tautomers are twice as high as $0.25$.
Moreover, the ionic transition states $S^\ddagger_\pm$ do not emerge
unlike the water-mediated case in which $p(S^\ddagger_\pm) = 0.25$.
In the meantime, the equilibrium probability of the neutral
transition state is still kept at the order of $10^{-30}$ at
physiological temperatures.

In this context, not only the conformational change induced in the
enzyme increases the likelihood of the formation of quantum H-bonds
with the nucleotide, but also the subsequent conformational change
that brings the enzyme back to its initial state makes quantum
H-bonds a more efficient resource for the tautomerization process.

\subsubsection{Classically correlated H-bonds}

The assumption of $\tau_2 < \tau_D$ isn't necessary to derive our
insight into the possible role of proton tunneling in an induced-fit
model of enzyme catalysis. Once the quantum correlations are created
by proton tunneling events through H-bonds, the enzyme can use them
as a resource to convert S to P by the same mechanism under
consideration even if these correlations rapidly decay to classical
correlations during decoherence.

In the case of $\tau_2 > \tau_D$, the final state of the
pseudo-spins of ES at the end of the binding stage should be a
classically correlated state rather than a four-qubit entangled
state:
\begin{eqnarray} \begin{aligned}
\rho_{1234}(t_3) \; &= \; a^2 |0011\rangle\langle0011| + b^2
|0101\rangle\langle0101| + c^2 |0110\rangle\langle0110| \\
&\; + d^2 |1001\rangle\langle1001| + e^2 |1010\rangle\langle1010| +
f^2 |1100\rangle\langle1100| .
\end{aligned} \end{eqnarray}

After the pseudo-spins of E are detached from the pseudo-spins of S
and returned back to their initial state by the post-selection
measurement induced by the second conformational change, the
nucleotide ends up in a mixture of S and P as below:
\begin{eqnarray} \begin{aligned} \label{State_ESin2}
\rho_{\text{NE}}(t_4) &= |G\rangle\langle G|_C \otimes \big(
{d^{\prime}}^2 |10\rangle \langle10|_{12} + {b^{\prime}}^2
|01\rangle \langle01|_{12}\big) \otimes |E\rangle \langle E| \\
&= \big({d^{\prime}}^2 |S\rangle\langle S| + {b^{\prime}}^2
|P\rangle\langle P| \big) \otimes |E\rangle \langle E| \\
&\equiv |ES^\ddagger\rangle \langle ES^\ddagger| ,
\end{aligned} \end{eqnarray}
where $b^\prime$ and $d^\prime$ are the re-normalized $b$ and $d$,
as before. If we feed master equation (\ref{MasterEq}) with this
incoherent superposition state instead of the coherent superposition
state $|\psi(t_4)\rangle_{\text{N}}$ given in (\ref{State_ESin1}),
the steady state solution $\rho_{\text{N}|cH\!B}^{eq}$ obtained is
the same density matrix with $\rho_{\text{N}|qH\!B}^{eq}$ in
(\ref{State_ESfin1}). Hence, the equilibrium probabilities remain
the same in both cases. Our conclusion still holds even if the
quantum correlations of H-bonds found in the reaction intermediate
ES are converted to classical correlations by strong decoherence
just at the beginning of the catalytic stage.

\subsubsection{Activation energy}

A state transformation from $\rho$ to $\rho^\prime$ is said to be
thermodynamically favorable only if the free energy $F$ goes down,
i.e., $\Delta F = F[\rho^\prime] - F[\rho] \leq 0$. Biological
systems can carry out thermodynamically unfavorable state
transformations by coupling them to favorable ones. However, even
thermodynamically favorable biochemical state transformations
generally involve the formation of transition states with higher
free energies than initial states. Enzymes lower these activation
barriers but are not expected to remove them completely. The
difference in the free energy between the transition and initial (or
final) states guarantees that this catalyzed transformation is
sufficiently slow for the organism inside which it takes place.

It is worth remembering that the tautomerization process of interest
is not assumed to be the ultimate catalytic function of the
multifunctional enzyme, but rather supports a secondary catalytic
function required for the ultimate one. In this respect, an
interesting question to ask is whether all enzyme-catalyzed
reactions have a rate-limiting step between the initial and final
states. In other words, can an enzyme provide the formation of a
transition state with a free energy lower than the free energy of
the initial (or final) state in specific cases? The simple model
investigated in this study points to a surprising possible answer to
this question in what follows.

The free energy of the nucleotide can be calculated by:
\begin{eqnarray}
\begin{aligned}
F[\rho] &= \langle E \rangle - R \, T S[\rho] = tr[\rho \, H_N] + R
\, T tr[\rho \, \log_2 \rho]
\end{aligned} \end{eqnarray}
where $R$ is the universal gas constant, $1.987$ cal/(mol K). As the
states of S, P, and S$^\ddagger$ are pure states, i.e., as
$tr[\rho^2] = 1$ for these tautomers, their von Neumann entropies
vanishes. These states are also eigenstates of $H_N$ which means
that average energies taken over them are nothing but the
corresponding eigenenergies. Hence, we end up with
\begin{equation}
F\big[|S\rangle \langle S|\big] = \epsilon_1 < F\big[|P\rangle
\langle P|\big] = \epsilon_2 < F\big[|S^\ddagger\rangle \langle
S^\ddagger |\big] = \epsilon_3.
\end{equation}

\begin{figure}[t]
    \centering
    \includegraphics[width=0.55 \linewidth]{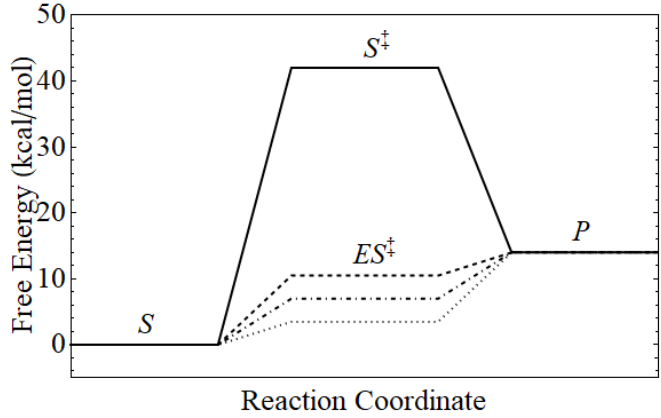}
    \caption{Alternative reaction pathways from S
    to P. Uncatalysed S (solid line) requires a high activation energy
    to reach S$^\ddagger$. Here,
    $E_e - E_g = 42$ kcal/mol, $\epsilon_+ - \epsilon_- = 14$ kcal/mol and
    $F\big[|S\rangle \langle S|\big]$ is set to zero
    as explained in the text. E is found to lower the activation energy
    by leading the reaction into an alternative chemical pathway (ES$^\ddagger$).
    Quantum correlations generated through H-bonds enable E to lower the activation energy
    down to a much lower value in-between the energies of S and P.
    $T = 37.5 ^\circ C$, $\tau_2 < \tau_D$ and ${b^\prime}^2$ is fixed to
    $3/4$, $2/4$ and $1/4$ respectively for the dashed, dot-dashed, and dotted lines.}
\label{Fig_En}
\end{figure}

We consider the free energy of the transition state ES$^\ddagger$
that appears after the binding stage of enzyme catalysis and study
the case in which quantum correlations that are generated through
the H-bonds of ES is preserved until the start of the catalytic
stage, i.e., $\tau_2 < \tau_D$. $|ES^\ddagger\rangle$ is separable
(see Eq. (\ref{State_ESin1})) and the reduced state of the
nucleotide equals $|ES^\ddagger\rangle_{\text{N}} = |G\rangle_C
\otimes (b^{\prime} |01\rangle + d^{\prime} |10\rangle)_{12}$ where
$d^\prime = \sqrt{1 - {b^\prime}^2}$. The expected value of the
energy over this state is $tr\big[|ES^\ddagger\rangle \langle
ES^\ddagger|_{\text{N}} \, H_N\big] = {b^{\prime}}^2 \epsilon_2 + (1
- {b^\prime}^2) \epsilon_1$. As it is a pure state, its von Neumann
entropy vanishes and its free energy becomes
\begin{equation}
F\big[|ES^\ddagger\rangle \langle ES^\ddagger|_{\text{N}}\big] =
{b^{\prime}}^2 \epsilon_2 + (1 - {b^\prime}^2) \epsilon_1 ,
\end{equation}
which corresponds to a free energy just in-between the free energy
of S and P, a value much lower than expected in biochemical
reactions (Fig. \ref{Fig_En}).

In the case of $\tau_2 \nless \tau_D$, $|ES^\ddagger\rangle$ is
separable again (see Eq. (\ref{State_ESin2})) but the reduced state
of the nucleotide equals $|ES^\ddagger\rangle \langle
ES^\ddagger|_{\text{N}} = |G\rangle\langle G|_C \otimes \big( (1 -
{b^\prime}^2) |10\rangle \langle10|_{12} + {b^{\prime}}^2 |01\rangle
\langle01|_{12}\big)$. As $tr\big[|ES^\ddagger\rangle \langle
ES^\ddagger|_{\text{N}} \, H_N\big]$ is also equal to
${b^{\prime}}^2 \epsilon_2 + (1 - {b^\prime}^2) \epsilon_1$, the
expected value of the energy between S and P along the reaction
coordinate is not affected by decoherence. However, unlike the
previous case, the reduced state of the nucleotide is not a pure
state, which means that its von Neumann entropy does not vanish:
$S\big[|ES^\ddagger\rangle \langle ES^\ddagger|_{\text{N}}\big] = -
{b^{\prime}}^2 \log_2 {b^{\prime}}^2 - (1 - {b^{\prime}}^2) \log_2(1
- {b^{\prime}}^2) \, > \, 0$. The free energy then drops further by
the loss of quantumness of correlations in the H-bonds as below:
\begin{equation}
F\big[|ES^\ddagger\rangle \langle ES^\ddagger|_{\text{N}}\big] =
{b^{\prime}}^2 \epsilon_2 + (1 - {b^\prime}^2) \epsilon_1 + R \, T
\big( {b^{\prime}}^2 \log_2 {b^{\prime}}^2 + (1 - {b^{\prime}}^2)
\log_2(1 - {b^{\prime}}^2) \big) .
\end{equation}

However, this entropy-based decrease is not likely to be significant
as $R \, T < 1$ kcal/mol at physiological temperatures and
$S\big[|ES^\ddagger\rangle \langle ES^\ddagger|_{\text{N}}\big] \,
\leq \, 1$. Thus, $F\big[|ES^\ddagger\rangle \langle
ES^\ddagger|_{\text{N}}\big]$ is expected to be in-between
$F\big[|S\rangle \langle S|\big]$ and $F\big[|P\rangle \langle
P|\big]$ once again.

We can numerically compare the activation energies in the catalyzed
and uncatalysed reactions using the model parameters that we chose
above (Fig. \ref{Fig_En}). Let's fix $F\big[|S\rangle \langle
S|\big]$ to zero for simplicity. To do so, both of the conformation
and pseudo-spin terms of $\epsilon_1$ are equated to zero, i.e.,
$E_g = 0$ and $W_1 + \lambda_{12} = 0$. Furthermore, we assume
without any loss of generality that $\big((W_1 - W_2)^2 + 4
J_{12}^2\big)^{1/2} \approx W_2 - W_1$. Then we end up with
$F\big[|P\rangle \langle P|\big] \approx 14$ kcal/mol and
$F\big[|S^\ddagger\rangle \langle S^\ddagger|\big] \approx 42$
kcal/mol. When we take $T = 37.5 ^\circ C$ and ${b^\prime}^2 = 1/2$
(or equivalently $b = d = 1/\sqrt{6}$) as before,
$F\big[|ES^\ddagger\rangle \langle ES^\ddagger|_{\text{N}}\big]$
becomes $7$ kcal/mol if $\tau_2 \le \tau_D$, and $6.38$ kcal/mol
otherwise. When ${b^\prime}^2$ is raised to $3/4$, these values
change to $10.5$ and $10$ kcal/mol respectively. On the other hand,
a decrease in ${b^\prime}^2$ to $1/4$ also lowers
$F\big[|ES^\ddagger\rangle \langle ES^\ddagger|_{\text{N}}\big]$ to
$3.5$ for $\tau_2 < \tau_D$ and $3$ kcal/mol for $\tau_2 > \tau_D$.
Hence the rate enhancement the enzyme brings about is in the range
of $22$ to $27$ orders of magnitude for $1/4 \leq {b^\prime}^2 \leq
3/4$ and $T = 37.5 ^\circ C$, based on the first-order rate equation
$k = \frac{k_B T}{h} e^{\frac{- \Delta F^\ddagger}{R \, T}}$, where
$k_B$ is the Boltzmann constant and $h$ is the Planck constant. The
lower limit of this rate enhancement can be decreased down to $5
\times 10^{19}$ by increasing the upper limit of ${b^\prime}^2$ up
to unity. Both limits can be pulled down by squeezing the energy
levels of the Hamiltonian given in (\ref{Ham_S}), e.g. when $1/4
\leq {b^\prime}^2 < 1$, $T = 37.5 ^\circ C$, $E_e - E_g = 34$
kcal/mol, and $\epsilon_+ - \epsilon_- = 19$ kcal/mol, the enzyme
leads to a rate enhancement in the range of $10$ to $20$ orders of
magnitude. Note that although this enhancement is more reasonable
when compared to the experimentally estimated upper limits, it does
not change our main proposition that the free energy of the
transition state is smaller than the free energy of the product in
the enzyme catalyzed reaction.

\section{Conclusions}

We showed that the quantum correlations generated through H-bonds
can be used as a resource in an induced-fit mechanism that leads
generic nucleotide tautomerization into an alternative chemical
pathway. This increases the occurrence of the minor tautomeric form
P even if the conversion of the major tautomeric form S to P isn't
possible in the absence of the enzyme. Moreover, the advantages of
this scenario go beyond that. As the new transition state that
emerges in this pathway corresponds to the quantum superposition of
S and P, the free energy for the new transition state is found to be
between the individual free energies of S and P. Hence, the enzyme
provides an alternative reaction pathway with a dramatically smaller
activation energy (Fig. \ref{Fig_En}). Moreover, such a reactive
pathway doesn't require anything more than a passive transformation
of the
enzyme's configuration in the catalytic stage.\\

\begin{acknowledgments}

O.P. thanks TUBITAK 2214-Program for financial support. T.F. and
V.V. thank the Oxford Martin Programme on Bio-Inspired Quantum
Technologies, the EPSRC and the Singapore Ministry of Education and
National Research Foundation for financial support.

\end{acknowledgments}


\begin{thebibliography}{99}

\bibitem{1999_Fersht} Fersht, A. 1999, \textit{Structure and mechanism in protein science:
a guide to enzyme catalysis and protein folding}, 2nd eds. W.H.
Freeman, New York.

\bibitem{2004_Science_ReviewOfES} Garcia-Viloca, M., Gao, J., Karplus, M., and Truhlar,
D.G., 2004, How enzymes work: analysis by modern rate theory and
computer simulations. \textit{Science} \textbf{303}, 186--195 (doi:
10.1126/science.1088172).

\bibitem{InducedFit_1958} Koshland, D. E., 1958,
Application of a theory of enzyme specificity to protein synthesis.
\textit{Proc. Natl. Acad. Sci. USA} \textbf{44}, 98--104 (doi:
10.1073/pnas.44.2.98).

\bibitem{InducedFit_2002} Hammes, G. G., 2002, Multiple conformational changes in enzyme
catalysis. \textit{Biochemistry} \textbf{41}, 8221--8228 (doi:
10.1021/bi0260839).

\bibitem{InducedFit_2007} Savir, Y. and
Tlusty, T., 2007, Conformational proofreading: the impact of
conformational changes on the specificity of molecular recognition.
\textit{PLoS ONE} \textbf{2}, e468 (doi:
10.1371/journal.pone.0000468).

\bibitem{ConfSelect_1999}
Ma, B., Kumar, S., Tsai, C. J.,and Nussinov, R., 1999, Folding
funnels and binding mechanisms. \textit{Protein Eng.} \textbf{12},
713--720 (doi: 10.1093/protein/12.9.713).

\bibitem{ConfSelect_2002}
Ma, B., Shatsky, M., Wolfson, H. J., and Nussinov, R., 2002,
Multiple diverse ligands binding at a single protein site: a matter
of pre-existing populations. \textit{Protein Sci.} \textbf{11},
184--197 (doi: 10.1110/ps.21302).

\bibitem{HTInE_1989__KIE} Cha, Y.,
Murray, C.J., and Klinman J.P., 1989, Hydrogen tunneling in enzyme
reactions. \textit{Science} \textbf{243}, 1325--1330 (doi:
10.1126/science.2646716).

\bibitem{HTInE_1996__KIE}
Nesheim, J. C. and Lipscomb, J. D., 1996, Large kinetic isotope
effects in methane oxidation catalyzed by methane monooxygenase:
evidence for C-H bond cleavage in a reaction cycle intermediate.
\textit{Biochemistry} \textbf{35}, 10240--10247 (doi:
10.1021/bi960596w).

\bibitem{HTInE_1998__KIE}
Whittaker, M. M., Ballou, D. P., and Whittaker, J. W., 1998, Kinetic
isotope effects as probes of the mechanism of galactose oxidase.
\textit{Biochemistry} \textbf{37}, 8426--8436 (doi:
10.1021/bi980328t).

\bibitem{HTInE_1999_1__KIE_Conf}
Basran, J., Sutcliffe, M. J., and Scrutton, N. S., 1999, Enzymatic
H-transfer requires vibration-driven extreme tunneling.
\textit{Biochemistry} \textbf{38}, 3218--3222 (doi:
10.1021/bi982719d).

\bibitem{HTInE_1999_2__KIE_Conf} Kohen, A., Cannio, R., Bartolucci,
S., Klinman, J.P., 1999, Enzyme dyamics and hydrogen tunneling in a
thermophilic alcohol dehydrogenase. \textit{Nature} \textbf{399},
496--499 (doi: 10.1038/20981).

\bibitem{HTInE_2000__KIE} Harris, R., Meskys, R., Sutcliffe, M.
J. and Scrutton, N. S., 2000, Kinetic studies of the mechanism of
carbon-hydrogen bond breakage by the heterotetrameric sarcosine
oxidase of \textit{Arthrobacter} sp. 1-IN. \textit{Biochemistry}
\textbf{39}, 1189--1198 (doi: 10.1021/bi991941v).

\bibitem{HTInE_2009__KIE}
Allemann, R.K. and Scrutton, N.S., 2009, \textit{Quantum tunnelling
in enzyme-catalysed reactions}. RSC, Cambridge (doi:
10.1039/9781847559975).

\bibitem{HTInE_2010__KIE} Sen, A. and Kohen,
A., 2010, Enzymatic tunneling and kinetic isotope effects: chemistry
at the crossroads. \textit{J. Phys. Org. Chem.} \textbf{23},
613--619 (doi: 10.1002/poc.1633).

\bibitem{HTInE_1991__MD_Conf}
Borgis, D. and Hynes, J. T., 1991, Molecular dynamics simulation for
a model nonadiabatic proton transfer reaction in solution.
\textit{J. Chem. Phys.} \textbf{94}, 3619--3628 (doi:
10.1063/1.459733).

\bibitem{HTInE_1997__MD_Conf}
Antoniou, D. and Schwartz, S.D., 1997, Large kinetic isotope effects
in enzymatic proton transfer and the role of substrate oscillations.
\textit{Proc. Natl. Acad. Sci. USA} \textbf{94}, 12360--12365.

\bibitem{HTInE_2001__MD_Conf}
Antoniou, D. and Schwartz, S.D., 2001, Internal enzyme motions as a
source of catalytic activity: rate-promoting vibrations and hydrogen
tunneling. \textit{J. Phys. Chem. B} \textbf{105}, 5553--5558 (doi:
10.1021/jp004547b).

\bibitem{HTInE_2006_1__QMMM_Conf} Masgrau, L., Roujeinikova, A.,
Johannissen, L.O., Hothi, P., Basran, J., Ranaghan, K.E.,
Mulholland, A.J., Sutcliffe, M.J., Scrutton, N.S., Leys, D., 2006,
Atomic description of an enzyme reaction dominated by proton
tunneling. \textit{Science} \textbf{312}, 237--241 (doi:
10.1126/science.1126002).

\bibitem{HTInE_2006_2__QMMM_Conf} Sutcliffe, M. J., Masgrau, L.,
Roujeinikova, A., Johannissen, K. E., Hothi, P., Basran, J.,
Ranaghan, K. E., Mulholland, A. J., Leys, D., and Scrutton, N. J.,
2006, Hydrogen tunnelling in enzyme-catalysed H-transfer reactions:
flavoprotein and quinoprotein systems. \textit{Phil. Trans. R. Soc.
B} \textbf{361}: 1375--1386 (doi: 10.1098/rstb.2006.1878).

\bibitem{HTInE_2006_3__QMMM_Conf}
Hammes-Schiffer, S. and Watney, J. B., 2006, Hydride transfer
catalysed by \textit{Escherichia coli} and \textit{Bacillus
subtilis} dihydrofolate reductase: coupled motions and distal
mutations. \textit{Phil. Trans. R. Soc. B} \textbf{361}, 1365--1373
(doi: 10.1098/rstb.2006.1869).

\bibitem{HTInE_2012__QMMM_Conf}
Glowacki, D. R., Harvey, J. N., and Mulholland, A. J., 2012, Taking
Ockham's razor to enzyme dynamics and catalysis. \textit{Nat. Chem.}
\textbf{4}, 169--176 (doi: 10.1038/nchem.1244).

\bibitem{HTInE_2013__QMMM_Conf}
Klinman, J. P. and Kohen, A., 2013, Hydrogen tunneling links protein
dynamics to enzyme catalysis. \textit{Annu. Rev. Biochem.}
\textbf{82}, 471--496 (doi: 10.1146/annurev-biochem-051710-133623).

\bibitem{Anders_2017_JPC_HTunnelingInSLO} Jevtic, S. and Anders, J.,
2017, A qualitative quantum rate model for hydrogen transfer in
soybean lipoxygenase. \textit{J. Chem. Phys.} \textbf{147}, 114108
(doi:10.1063/1.4998941).

\bibitem{NoEffectOfT_2003}
Doll, K.M., Bender, B.R., Finke, R.G., 2003, The first experimental
test of the hypothesis that enzymes have evolved to enhance hydrogen
tunneling. \textit{J. Am. Chem. Soc.} \textbf{125}, 10877--10884
(doi: 10.1021/ja030120h).

\bibitem{NoEffectOfT_1991}
Hwang, J.K., Chu, Z.T., Yadav, A., Warshel, A. 1991, Simulations of
quantum-mechanical corrections for rate constants of
hydride-transfer reactions in enzymes and solutions. \textit{J.
Phys. Chem.} \textbf{95}, 8445--8448 (doi: 10.1021/j100175a009).

\bibitem{NoEffectOfT_1996}
Hwang, J.K. and Warshel, A., 1996, How important are quantum
mechanical nuclear motions in enzyme catalysis? \textit{J. Am. Chem.
Soc.} \textbf{118}, 11745--11751 (doi: 10.1021/ja962007f).

\bibitem{NoEffectOfT_2004}
Olsson, M.H., Siegbahn, P.E., Warshel, A., 2004, Simulations of the
large kinetic isotope effect and the temperature dependence of the
hydrogen atom transfer in lipoxygenase. \textit{J. Am. Chem. Soc.}
\textbf{126}, 2820--2828 (doi: 10.1021/ja037233l).

\bibitem{NoEffectOfT_2006}
Olsson, M.H., Marvi, J., Warshel, A., 2006, Transition state theory
can be used in studies of enzyme catalysis: lessons from simulations
of tunnelling and dynamical effects in lipoxygenase and other
systems. \textit{Phil. Trans. R. Soc. B} \textbf{361}, 1417--1432
(doi: 10.1098/rstb.2006.1880).

\bibitem{NoEffectOfT_2010} Williams, I.H.,
2010, Quantum catalysis? A comment on tunnelling contributions for
catalysed and uncatalysed reactions. \textit{J. Phys. Org. Chem.}
\textbf{23}, 685--689 (doi: 10.1002/poc.1658).

\bibitem{2010_McKenzie} Bothma, J. P., Gilmore, J. B., and
McKenzie, R. H., 2010, The role of quantum effects in proton
transfer reactions in enzymes: quantum tunneling in a noisy
environment? \textit{New J. Phys.} \textbf{12}, 055002 (doi:
10.1088/1367-2630/12/5/055002).

\bibitem{YesOrNo_2004} Ball, P., 2004, Enzymes: By chance, or by
design? \textit{Nature} \textbf{431}, 396--397 (doi:
10.1038/431396a).

\bibitem{YesOrNo_2006}
Dutton, P. L., Munro, A. W., Scrutton, N. J., and Sutcliffe, M. J.,
2006, Introduction. Quantum catalysis in enzymes: beyond the
transition state theory paradigm. \textit{Phil. Trans. R. Soc. B}
\textbf{361}: 1293--1294 (doi: 10.1098/rstb.2006.1879).

\bibitem{YesOrNo_2017} Brookes, J. C., 2017, Quantum
effects in biology: golden rule in enzymes, olfaction,
photosynthesis and magnetodetection. \textit{Proc. R. Soc. A}
\textbf{473}, 20160822 (doi: 10.1098/rspa.2016.0822).

\bibitem{2008_TautomeraseSF} Poelarends, G. J., Veetil, V. P. and Whitman, C.
P., 2008, The chemical versatility of the beta-alpha-beta fold:
catalytic promiscuity and divergent evolution in the tautomerase
superfamily. \textit{Cell. Mol. Life Sci.} \textbf{65}, 3606--3618
(doi: 10.1007/s00018-008-8285-x).

\bibitem{2016_TautoRequiredByE} Zhu, C., Lu, L., Zhang, J., Yue, Z.,
Song, J., Zong, S., Liu, M., Stovicek, O., Gao, Y. Q. and Yi, C.,
2016, Tautomerization-dependent recognition and excision of
oxidation damage in base-excision DNA repair. \textit{Proc. Natl.
Acad. Sci. U S A} \textbf{113}, 7792--7797 (doi:
10.1073/pnas.1604591113).

\bibitem{2015_TautomeraseRoleInRNA} Singh, V., Fedeles, B. I., and
Essigmann, J. M., 2015, Role of tautomerism in RNA biochemistry.
\textit{RNA} \textbf{21}, 1--13 (doi: 10.1261/rna.048371.114).

\bibitem{InducedFit_2013_DNA}, Zahurancik, W. C., Klein, S. J.
and Suo, Z., 2013, Kinetic mechanism of DNA polymerization catalyzed
by human DNA polymerase $\varepsilon$. \textit{Biochemistry}
\textbf{52}, 7041--7049 (doi: 10.1021/bi400803v).

\bibitem{InducedFit_2014_DNA} Miller, B.R. III, Parish, C. A. and Wu, E.
Y., 2014, Molecular dynamics study of the opening mechanism for DNA
polymerase I. \textit{PLoS Comput. Biol.} \textbf{10}, e1003961
(doi: 10.1371/journal.pcbi.1003961).

\bibitem{InducedFit_2015_DNA-1} Moscato, B., Swain, M. and Loria, J. P.,
2015, Induced fit in the selection of correct versus incorrect
nucleotides by DNA polymerase $\beta$. \textit{Biochemistry}
\textbf{55}, 382--395 (doi: 10.1021/acs.biochem.5b01213).

\bibitem{InducedFit_2015_DNA-2}, Evans, G. W., Hohlbein, J., Craggs, T.,
Aigrain, L. and Kapanidis, A. N., 2015, Real-time single-molecule
studies of the motions of DNA polymerase fingers illuminate DNA
synthesis mechanisms. \textit{Nucleic Acids Res.} \textbf{43},
5998--6008 (doi: 10.1093/nar/gkv547).

\bibitem{InducedFit_2016_DNA} Batra, V. K., Beard, W. A., Pedersen, L. C.
and Wilson, S. H., 2016, Switch to standard view structures of DNA
polymerase mispaired DNA termini transitioning to pre-catalytic
complexes support an induced-fit fidelity mechanism.
\textit{Structure} \textbf{24}, 1863--1875 (doi:
10.1016/j.str.2016.08.006).

\bibitem{2012_McKenzie}
McKenzie, R. H., 2012, A diabatic state model for donor-hydrogen
vibrational frequency shifts in hydrogen bonded complexes.
\textit{Chem. Phys. Lett.} \textbf{535}, 196--200 (doi:
10.1016/j.cplett.2012.03.064).

\bibitem{2014_McKenzie} McKenzie, R. H., Bekker, C., Athokpam, B.,
and Ramesh, S. G., 2014, Effect of quantum nuclear motion on
hydrogen bonding. \textit{J. Chem. Phys.} \textbf{140}, 174508 (doi:
10.1063/1.4873352).

\bibitem{Anders_2011_Chaos_ComplexityAndquantum} Anders, J. and
Wiesner, K., 2011, Increasing complexity with quantum physics.
\textit{Chaos} \textbf{21}, 037102 (doi = 10.1063/1.3640753).

\bibitem
{BreuerAndPetruccione-2002} Breuer, H. P. and Petruccione F., 2002,
\textit{The theory of open quantum systems}. pp. 130 - 137 (New
York: Oxford University Press)

\bibitem{Kosloff_2013_MasterEqsInQTD} Kosloff, R., 2013, Quantum Thermodynamics: A Dynamical
Viewpoint. \textit{Entropy} \textbf{15}, 2100--2128 (doi:
10.3390/e15062100).

\bibitem{Parameters_T} Fan, J. C. Shang, Z. C., Liang, J., Liu, X. H.,
Jin, H., 2010, Systematic theoretical investigations on the
tautomers of thymine in gas phase and solution. \textit{J. Mol.
Struct. THEOCHEM} \textbf{939}, 106--111 (doi:
10.1016/j.theochem.2009.09.047).

\bibitem{Parameters_G} Li, H. F., Zhang, L. S., Fan, X. L., 2014,
Metal counterion modulated single proton transfer process in guanine
base. \textit{Comput. Theo. Chem.} \textbf{1032}, 90--96 (doi:
10.1016/j.comptc.2014.01.026).

\bibitem{Parameters_A} Li, H., Zhang, L., Zhou, H., Wang, Y.,
Fan, X., 2015, Theoretical studies on the single proton transfer
process in adenine base. \textit{J. Phys. Org. Chem.} \textbf{28},
755--760 (doi: 10.1002/poc.3479).


\end{thebibliography}
\end{document}